\documentclass[aps,pra,twocolumn,superscriptaddress]{revtex4-1}

\usepackage{graphicx}
\usepackage{amssymb}
\usepackage{amsmath}
\usepackage{amsfonts}
\usepackage{braket}
\usepackage{color}
\usepackage[utf8]{inputenc}		
\usepackage[english]{babel}

\begin{document}

\title{
Magnon topological transition in skyrmion crystal 
} 
\author{V.~E. Timofeev}
\email{viktor.timofeev@spbu.ru}
\affiliation{NRC ``Kurchatov Institute'', Petersburg Nuclear Physics Institute, Gatchina
188300, Russia}
\affiliation{St.Petersburg State University, 7/9 Universitetskaya nab., 199034
St.~Petersburg, Russia} 
\author{Yu.~V. Baramygina}
\affiliation{NRC ``Kurchatov Institute'', Petersburg Nuclear Physics Institute, Gatchina
188300, Russia}
\affiliation{St.Petersburg State University, 7/9 Universitetskaya nab., 199034
St.~Petersburg, Russia} 
\author{D.~N. Aristov}
\affiliation{NRC ``Kurchatov Institute'', Petersburg Nuclear Physics Institute, Gatchina
188300, Russia}
\affiliation{St.Petersburg State University, 7/9 Universitetskaya nab., 199034
St.~Petersburg, Russia}  
 
\begin{abstract} 
We study the magnon spectrum  in skyrmion crystal formed in thin ferromagnetic films with Dzyalosinskii-Moria interaction  in presence of  magnetic field. Focusing on two low-lying observable magnon modes and employing stereographic projection method, we develop a theory demonstrating a topological transition in the spectrum. Upon the increase of magnetic field, the gap between two magnon bands closes, with the ensuing change in the topological character of both bands. This phenomenon of gap closing, if confirmed in magnetic resonance experiments, may deserve  further investigation by thermal Hall conductivity experiments. 
\end{abstract}

\maketitle

%
{\bf Introduction}.
Magnetic skyrmions are topologically nontrivial whirls of local magnetization. 
Nowadays they are discussed in  context of  development of novel memory devices  \cite{Vakili_2021} and unconventional computing \cite{Lee2022}.
Besides the practical interest, the topic of magnetic skyrmions raises also a number of questions of fundamental nature.

One direction of active research concerns isolated metastable skyrmions with long lifetime \cite{fert2017magnetic}. However, in a wide range of materials the magnetic skyrmions are ordered into lattices \cite{everschor2018perspective}, co-called skyrmion crystals (SkX) \cite{Nagaosa2013}, and it opens another interesting direction for applications in magnonics \cite{garst2017collective}.

The magnetic skyrmions in thin ferromagnetic films were discovered by Belavin and Polyakov as metastable configurations of local magnetization \cite{Belavin1975}. Since then different mechanisms of skyrmion stabilization were proposed, and Dzyaloshinskii-Moriya interaction (DMI) is  one of them \cite{bogdanov1989thermodynamically,BOGDANOV1994255}. DMI appears in non-centrosymmetric  chiral magnets, e.g.\ in B20-type compound  MnSi, where magnetic skyrmions in a form of SkX were observed for the first time \cite{muhlbauer2009skyrmion}.

The internal dynamics of isolated skyrmions is quite involved on its own \cite{schutte2014magnon,lin2014internal}, and the ordering of skyrmions into lattices leads to even more complicated band structure of SkX excitations \cite{roldan2016,garst2017collective,Timofeev2022}. 
Besides the topologically trivial Goldstone mode \cite{petrova2011, Timofeev2023a},  other low-energy bands may also possess non-zero Chern numbers \cite{roldan2016}.
It is known that  nontrivial topology of magnon bands may give rise to the edge states \cite{roldan2016,Diaz2020, Maeland2022}, 
%
%
or thermal Hall transport \cite{Hoogdalem2013,PhysRevB.89.054420}.

The band structure of low-lying magnons was investigated using the stereographic projection approach \cite{Timofeev2022}. 
A special approach for consideration of the gyrotropic mode of SkX was recently devised in \cite{Timofeev2023a}. 
The latter approach is generalized in the present work to include two modes of higher energy, which are observable in magnetic resonance experiments.  
In accordance with earlier numerical findings  \cite{Diaz2020}, we show a topological transition happening at some intermediate field value inside the stability region of SkX phase.

%
%
{\bf Model.}
We consider a model of planar ferromagnet in presence of DMI and uniform external magnetic field perpendicular to the plane. The energy density is given by:
\begin{equation}
\mathcal{E} =   \frac{C}{2}  \partial_{\mu}S_{i}\partial_{\mu}S_{i} - 
D\epsilon_{\mu ij} S_{i}\partial_{\mu}S_{j}  - B  S_{3},
\label{classicalenergy}
\end{equation}
with $C$  exchange parameter, $D$ is   DMI constant, and $B$ the magnetic field. It is convenient to define the unit length as $l=C/D$, and to measure energy density in  units of $ CS^2 l^{-2} = S^2D^2/C$. The energy, $\mathcal{E}$,  then depends only on the dimensionless parameter $b=BC /SD^2$. In the low temperature limit considered here the local magnetization is saturated to its maximum value, $S$, and  $\mathbf{S} = S \mathbf{n}$, with $|\mathbf{n}|=1$.  This planar model is  also applicable to thin films, as long as its  thickness does not exceed $l$. 
The magnetic dipolar interaction is ignored here, because in thin film geometry it is reduced to uniaxial anisotropy and may lead to only minor changes of SkX parameters.

To study dynamics of local magnetization, we  use the Lagrangian, $ \mathcal{L} = \mathcal{T} - \mathcal{E}$,  with the  kinetic term \cite{D_ring_1948}:
\begin{equation}
\mathcal{T}=
\frac { S}{ \gamma_0}(1-\cos{\theta})\dot{\varphi}\,,
\label{eq:kinetic}
\end{equation}
here  $\varphi$ and $\theta$ are the angles of the magnetization direction $\mathbf{n}=(\cos\varphi\sin\theta,\sin\varphi\sin\theta,\cos\theta)$, and $\gamma_0$ is gyromagnetic ratio. The above form of the Lagrangian results in the well known Landau-Lifshitz equation.  In what follows, we include the factor $S/ \gamma_0$  into the time scale. 

We use the stereographic projection method, by representing $\mathbf{n}$ as follows 
\begin{equation}
n_1 + i \, n_2  = \frac{2f}{1 + f\bar{f}}\,,\quad 
n_3 = \frac{1 - f\bar{f}}{1 + f\bar{f}},
\label{eq:stereo}
\end{equation}
with $f$ a complex-valued function, and $\bar{f}$ its complex conjugate. 
We suggest to describe a single skyrmion by its stereographic function in the form 
\begin{equation}
f_1=\frac{i \, z_0 \,\kappa(z\bar{z}/z_0^2)}{\bar{z}},
\label{eq:anz}
\end{equation}
with $z=x+iy$, $\kappa$ a smooth real-valued profile function, and $z_0$ is a skyrmion size parameter. In case of one skyrmion  Eq.\  \eqref{eq:anz}  
reproduces the profile obtained by usual bubble domain ansatz, while the ansatz \eqref{eq:anz} is more convenient for the  description of SkX profile.  It was shown previously that the multi-skyrmion configuration can be constructed as a sum of stereographic functions of individual skyrmions \cite{Timofeev2019}. Particularly, we propose to model the regularly arranged  hexagonal SkX by the following stereographic function:
\begin{equation}
f_{SkX} = \sum\limits_{n,m} f_1 (\mathbf{r} - n \mathbf{a}_1 - m \mathbf{a}_2),
\label{SkXf0}
\end{equation}
where $\mathbf{a}_1 = (0,a)$, $\mathbf{a}_2 = (-\sqrt{3}a/2,a/2)$, and $a$ is a cell parameter of SkX. The static properties of this ansatz \eqref{eq:anz}-\eqref{SkXf0} were discussed previously in \cite{Timofeev2019,Timofeev2021}. It was shown that the proposed SkX configuration has lower energy than helix or uniform configuration at magnetic fields, $0.25 \alt b \alt 0.8$.
 
In terms of stereographic projection the equation of motion for $f(t)$ is highly nonlinear and cannot  be solved in general case. In previous works \cite{Timofeev2022,Timofeev_2023} we have discussed the normal modes of infinitesimal fluctuations of the function $f$. 
Recently we also developed a special approach for consideration of the gyrotropic mode of SkX \cite{Timofeev2023a}.
In the present work we generalize the latter approach for the analysis of two higher-energy modes, which are in principle experimentally observable.


{\bf The gyrotropic mode.}
To illustrate our specially devised approach, let us first remind how it was done for the gyrotropic mode. 
We assume that the skyrmion lattice is imperfect, so that  the position of the center of skyrmions may vary, whereas the individual images of skyrmions in the sum \eqref{SkXf0} are  unchanged. We write 
\begin{equation}
f_{SkX}  = \sum\limits_{l} f_1 (\mathbf{r} - n \mathbf{a}_1 - m \mathbf{a}_2 
+  \mathbf{u}_l) \,, 
\label{SkXf}
\end{equation}
and in the lowest order of $u_l$ we obtain
\begin{equation}
    f_{SkX}  = f_{0} +  \sum\limits_{l} \mathbf{u}_l \nabla f_1 (\mathbf{r} - \mathbf{r}^{(0)}_l),
\end{equation}
here $\mathbf{r}^{(0)}_l = n \mathbf{a}_1 - m \mathbf{a}_2$ and unperturbed $f_{0}$ is given by \eqref{SkXf0}. 

To be able to use our previous formulas  for the dynamics in \cite{Timofeev2022}, we pass to the variable    $\psi(\mathbf{r} )$, according to 
\begin{equation}
    \sum\limits_{l} \mathbf{u}_l \nabla f_1 (\mathbf{r} - \mathbf{r}^{(0)}_l)
  =   (1+f_0 \bar f_0)\, \psi(\mathbf{r} )   ,
\end{equation}
Introducing shorthand notation $f_j =  f_1 (\mathbf{r} - \mathbf{r}^{(0)}_j)$  we arrive to 
\begin{equation}
\begin{aligned}
         \begin{pmatrix}
           \psi \\ \bar \psi
       \end{pmatrix} 
       &= \frac1{1+f_0 \bar f_0} \sum_j \begin{pmatrix}
      \partial_{\bar z} f_j ,& \partial_{ z} f_j \\ \partial_{\bar z} \bar f_j  ,& \partial_{ z} \bar f_j 
    \end{pmatrix}  
    \begin{pmatrix}
         u^-_j \\ u^+_j
       \end{pmatrix} 
      \\ 
    & \equiv  \sum_j {\cal O}_j \begin{pmatrix}
         u^-_j \\ u^+_j
       \end{pmatrix}  \,,
\end{aligned}
\label{zeromode}
\end{equation} 
here $\mathbf{u}_j \nabla = u^+_j \partial_{z}  + u^-_j \partial_{\bar{z}}$ with 
 $\partial_{z} = (\partial_{x}-i\partial_{y})/2$ , $\partial_{\bar{z}} = (\partial_{x}+i\partial_{y})/2$ and $u^\pm_j = u^x_j \pm i u^y_j$, 
and $\bar \psi$ is complex conjugate of $ \psi$. 
 
The quadratic in displacements part of the Lagrangian attains then the form  
\begin{equation}
\begin{aligned}
\mathcal{L} & =\frac12 
 \sum_{lj} \begin{pmatrix}
         u^+_l ,&  u^-_l
       \end{pmatrix} \left( -i \hat{\mathcal K}_{lj}  \partial_t  
       -\hat{\mathcal{H}}_{lj} \right) 
       \begin{pmatrix}
         u^-_j \\   u^+_j
       \end{pmatrix} 
      \, ,  \\ 
      \hat{\mathcal K}_{lj} & = 
      \int d\mathbf{r}\, {\cal O}^\dagger_l  . \sigma_3 .  {\cal O}_j 
      \, ,  \\ 
        \hat{\mathcal H}_{lj} & =  
     \int d \mathbf{r}\, {\cal O}^\dagger_l .   \begin{pmatrix}
  (-i\nabla + \mathbf{A})^2 + U&
   V\\
  V^*&
   (i\nabla + \mathbf{A})^2 + U
\end{pmatrix} .{\cal O}_j  \,, 
\end{aligned} 
\label{eq:Lagr2}
\end{equation}
 where the explicit form of  $U$, $V$ and $\mathbf{A} $ is given in \cite{Timofeev2022}. 
  Due to rapid decrease of  $f_j =  f_1 (\mathbf{r} - \mathbf{r}^{(0)}_j)$ away from 
 the position of $j$th skyrmion, the quantities  $\hat{\mathcal K}_{lj} $ and $\hat{\mathcal H}_{lj}$ are reduced in fact to nearest-neighbors couplings. Further details about the dynamics of the gyrotropic mode can be found in \cite{Timofeev2023a}.
 
 %
%

{\bf Two higher-energy modes.}
Consider now two modes, which are observable in magnetic resonance experiments \cite{Timofeev_2023}, the so-called breathing(Br) and counter-clockwise modes(CCW). 
Instead of shifting the position of individual skyrmions,  we allow now to change their radius or phase. This type of variation corresponds to breathing mode.   In the simplified form of our ansatz we assume the shape in the form $f_1(\mathbf{r} )= iz_0/\bar {z}\exp(-c |z/z_0|^2 )$, with $c$ dependent on the field value.
The change in real-valued $z_0\to z_0 (1+\epsilon_1)$ may be written as 
$f_1(\mathbf{r} ) \to f_1(\mathbf{r} ) + \epsilon_1 z_0 \partial  f_1(\mathbf{r} )/\partial z_0 = 
 f_1(\mathbf{r} ) - \epsilon_1( z \partial  f_1(\mathbf{r} )/\partial z + \bar{z} \partial  f_1(\mathbf{r} )/\partial \bar{z} ) $. 
 Similarly, the infinitesimal change in phase $f_1(\mathbf{r} ) \to e^{i\epsilon_2} f_1(\mathbf{r} )$ can be written as 
$f_1(\mathbf{r} ) \to  f_1(\mathbf{r} ) +i \epsilon_2( z \partial  f_1(\mathbf{r} )/\partial z - \bar{z} \partial  f_1(\mathbf{r} )/\partial \bar{z} ) $.
Combining these formulas, and defining the complex-valued  $u_{br}=\epsilon_1-i\epsilon_2$, we obtain the first-order  variation in the form 
\[ u_{br} \, z\, \partial  f_1(\mathbf{r})/\partial z + \bar{u}_{br}\, \bar{z} \,\partial  f_1(\mathbf{r})/\partial \bar{z} \,.\]
We see here that the breathing mode has the same angular character at the origin as $f_1$, i.e.\  it behaves as  $e^{i\phi}$ at $r\to 0$. It should be contrasted to the  translational mode above, which behaves as $e^{2i\phi}$, i.e. has additional clockwise rotation, $ e^{i\phi}$, with respect to $f_1\propto e^{i\phi}$.
It is easy to show that another observable mode, the counter-clockwise rotation, is obtained by taking the variation 
\[ u_{ccw} \, z^2 \partial  f_1(r)/\partial z + \bar{u}_{ccw}\,\bar{z}^2 \partial  f_1(r)/\partial \bar{z} \,. \]
 
Further modeling is straightforward, and is done largely repeating the above steps for zero mode. 
First, we introduce the amplitudes referring to individual skyrmions, writing, e.g., for the breathing mode of the skyrmion centered at $\mathbf{r}^{(0)}_j$ : 
\[ u_{br,j}  (z-z_j^{(0)}) \frac{\partial}{\partial z}  f_j + 
\bar{u}_{br,j}(\bar{z}-z_j^{(0)}) \frac{\partial}{\partial \bar{z} } f_j  \,,
\]
and similarly for  $u_{ccw,j}$. 
Then we perform the integration over $\mathbf{r} $, obtain the Lagrangian on the lattice, in terms of the modes $u_{br,j}$, $u_{ccw,j}$ and pass to Fourier transform,  
$     u_{br,j}  = \sum _\mathbf{q} e^{i \mathbf{q} \mathbf{r}_j} u_{br,\mathbf{q}}$, and similarly for $u_{ccw,\mathbf{q}}$.  

It turns out that  the energies of the modes  $u_{br}$ and $u_{ccw}$ are  close and even coinciding  at some $b$. Therefore we  need to take into account  the hybridization of these two modes. 
We introduce a ``Dirac spinor'' object,  $\Psi^\dagger_ {\mathbf{q}} = \begin{pmatrix} 
\bar {u}_{br,\mathbf{q}} ,&  u_{br,\mathbf{q}}   ,&   \bar {u}_{ccw,\mathbf{q}} ,&  u_{ccw,\mathbf{q}} 
 \end{pmatrix}$. 
After some calculation we obtain the  effective Lagrangian for these two bands in the form 
\begin{equation}     \begin{aligned} 
   \mathcal{L}   &=    \frac12  \sum _\mathbf{q} 
   \Psi^\dagger_ {-\mathbf{q}} 
  \left( -i \hat{\mathcal K}_{\mathbf{q}}  \partial_t         -\hat{\mathcal{H}}_{\mathbf{q}} \right) 
  \Psi_ {\mathbf{q}} 
      \, ,  \\  
      \hat{\mathcal K}_{\mathbf{q}}  &=   
      \begin{pmatrix}
      1+ k_2 \gamma_s   , & 0, & k_a  \gamma_p ^\ast   , &  k_b  \gamma_p  \\
      0, &- 1- k_2  \gamma_s , &   k_b  \gamma_p ^\ast , &  k_a  \gamma_p \\
        k_a  \gamma_p  , &  k_b \gamma_p, & 1+ k_3 \gamma_s , & 0   \\
          k_b  \gamma_p ^\ast , &  k_a \gamma_p ^\ast ,& 0, &- 1- k_3 \gamma_s        
      \end{pmatrix}
       \,, \\ 
     \hat{\mathcal H}_{\mathbf{q}} &= \begin{pmatrix}
    p_1 + t_1  \gamma_s , & p_2 + t_2   \gamma_s  , & 
     - t_a \gamma_p ^\ast  , & t_b \gamma_p    \\  
      p_2 + t_2   \gamma_s  , &p_1 + t_1 \gamma_s , &
      - t_b  \gamma_p ^\ast , &  t_a  \gamma_p     \\    
     - t_a \gamma_p  , & - t_b  \gamma_p      , &     
     p_3 + t_3 \gamma_s , &  t_4   \gamma_d   \\  
      t_b  \gamma_p ^\ast , &  t_a \gamma_p ^\ast  , & 
     t_4   \gamma_d ^\ast   , &p_3 + t_3   \gamma_s   \\          
      \end{pmatrix} \,,    
    \end{aligned}
    \label{eq:L35}
\end{equation}
where  the sums over six nearest neighbors (NN) are  defined as 
  \begin{equation}     \begin{aligned}
    \gamma_s(\mathbf{q})  &=  \sum  _\mathbf{d} e^{-i \mathbf{q} \mathbf{d}} - 6  
    \\ & = 
    2\left( 2  \cos \tfrac{ \sqrt{3}} 2   q_x \cos \tfrac12  {q_y}   +  \cos q_y  -3  \right)
     \, , \\ 
    \gamma_p(\mathbf{q})  
    &=   \sum _\mathbf{d} e^{-i \mathbf{q} \mathbf{d}} e^{ i\phi_d} 
    \\  & = 
    2 \left(   \cos \tfrac{ \sqrt{3}} 2   q_x \sin \tfrac12  {q_y}   +  \sin q_y 
    \right. \\ &  \left.
    -   i   \sqrt{3}  \sin \tfrac{ \sqrt{3}} 2   q_x \cos \tfrac12  {q_y} \right) 
    \, , \\ 
    \gamma_d(\mathbf{q}) 
    &=   \sum _\mathbf{d} e^{-i \mathbf{q} \mathbf{d}} e^{ 2i\phi_d} 
    \\  & = 
    2 \left(   \cos \tfrac{ \sqrt{3}} 2   q_x \cos \tfrac12  {q_y}   -  \cos q_y 
    \right. \\ &  \left.
    -   i   \sqrt{3}  \sin \tfrac{ \sqrt{3}} 2   q_x \sin \tfrac12  {q_y} \right)
     \,,   
    \end{aligned}   
\end{equation}
with the property  $\gamma_{s,p,d}(0)  = 0$.  
Here we put the SkX cell parameter, $a$, to unity for simplicity of notation.  For optimal configuration $a$ is dependent on the field, $b$.  

\begin{figure}[t]
\center{\includegraphics[width=0.99\linewidth]{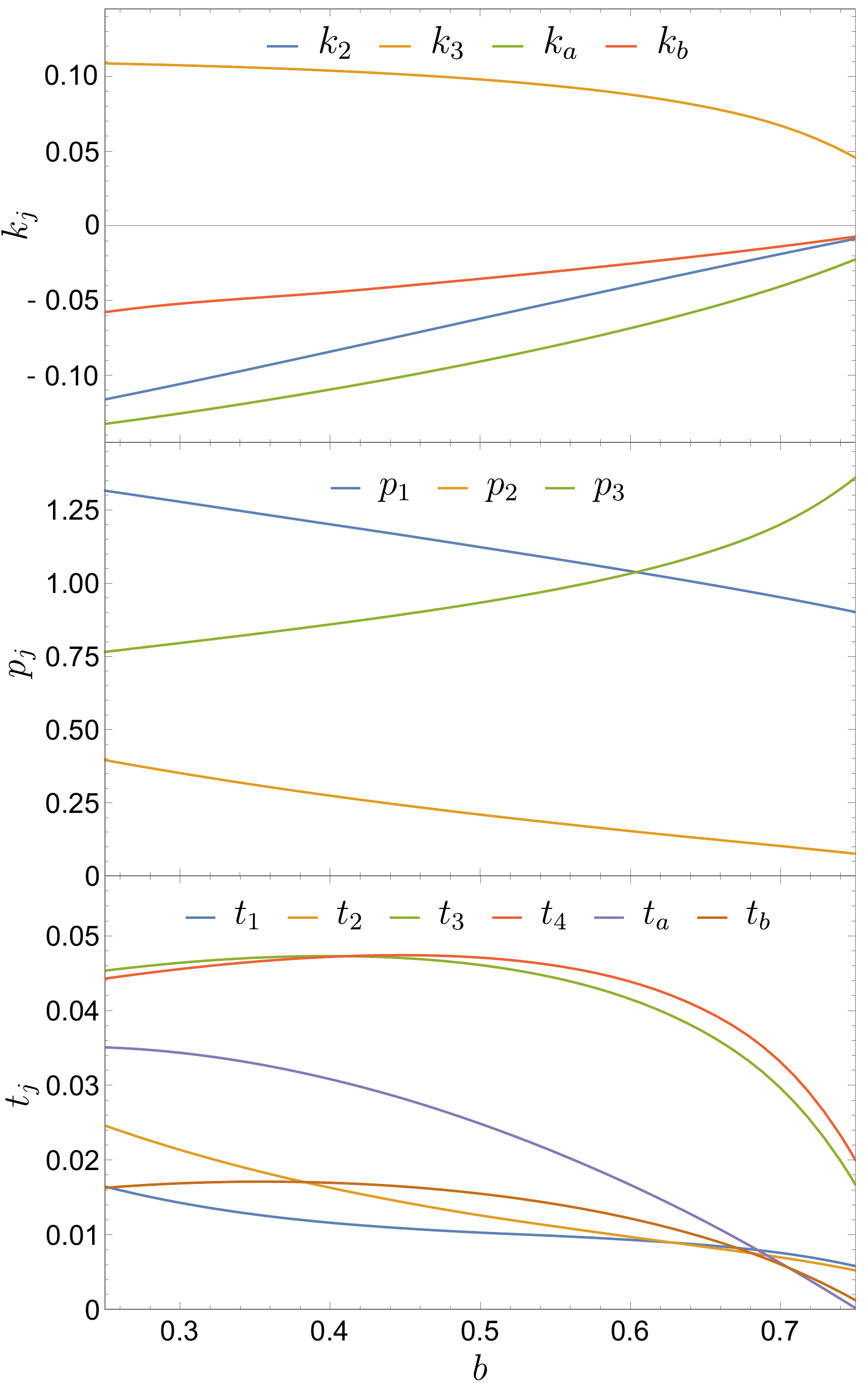}} 
\caption{Parameters of the Lagrangian \eqref{eq:L35}, describing breathing and counterclockwise modes as a function of the external magnetic field, $b$.}
\label{fig:param}
\end{figure}

The convenient property  $\hat{\mathcal K}_{\mathbf{q} =0 } = \mbox{diag }(1,-1,1,-1)$ was actually obtained by rescaling $u_{br,j}\to c_{br} u_{br,j}$ and $u_{ccw,j} \to c_{ccw}  u_{ccw,j} $ for all $j$ with appropriately chosen (and field dependent) $c_{br}$, $c_{ccw}$.  Numerically,  we find that coefficients $k_2<0< k_3$, $k_{a,b}<0$ are rather small  and their dependence on the field $b$ is shown in Fig. \ref{fig:param}. In this plot we also show other parameters, $p_i$, $t_i$, referring to potential part of the Lagrangian. 

Despite the relatively small value of $k_{3}$, one should be cautious when considering the vicinity  
of  $K$ point in the Brillouin zone, i.e., near $\mathbf{q}= \tfrac{2\pi}{3}(\sqrt{3},1)$. At this point the factor $\gamma_s$ has its minimum, $\gamma_s=-9$, so that  the value of $\hat{\mathcal K}_{\mathbf{q},33} $ becomes numerically small $1+ k_3 \gamma_s \simeq 0.07$ at low fields, $b\leq 0.4$. Still the value $\hat{\mathcal K}_{\mathbf{q},33} $ remains  positive for all $b$, as should be in well-defined theory. 

\begin{figure}[t]
\center{\includegraphics[width=0.99\linewidth]{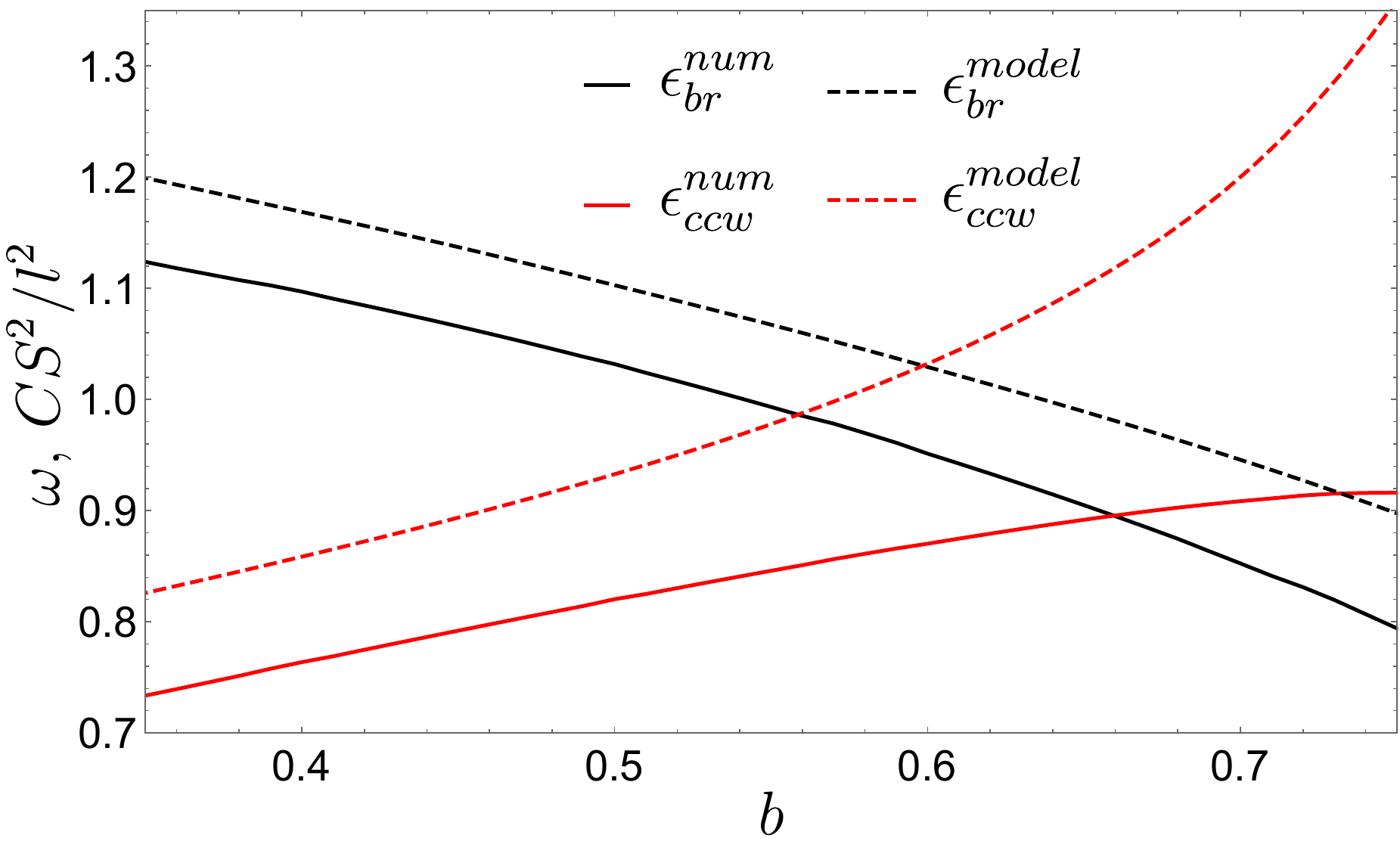}} 
\caption{ Energies of the breathing and counterclockwise modes exactly at the $\Gamma$ point, $\mathbf{q}=0$, calculated in two approaches. The results for effective model, Eq. \eqref{eqGamma}, are shown by the dashed lines, the results according to full-scale calculation, described in Ref.\  [\onlinecite{Timofeev2022}], are shown by solid lines. }
\label{fig:Gamma}
\end{figure}

The spectrum is  determined by the  equation 
$\mbox{det} (\hat{\mathcal L}_{\mathbf{q}} ) =0$, where $\hat{\mathcal L}_{\mathbf{q}} =\omega \hat{\mathcal K}_{\mathbf{q}}  -\hat{\mathcal{H}}_{\mathbf{q}}$.  The roots of this equation come in pairs, $\omega = \pm \epsilon_1(\mathbf{q})$  and $\omega = \pm \epsilon_2(\mathbf{q})$.
Exactly at the $\Gamma$ point, $ \mathbf{q} =0$, this equation can be easily analyzed. 
We have in this case  $\gamma_s = \gamma_p = \gamma_d=0$, and the matrix of Lagrangian becomes 
\begin{equation}
  \hat{\mathcal L}_{\mathbf{q}=0} =
  \begin{pmatrix}
        \omega -  p_1, &  -p_2, & 0, & 0 \\ 
        - p_2, & -\omega -  p_1, &   0, & 0 \\ 
        0, & 0, &  \omega -  p_3, &  0 \\ 
        0, & 0, &  0 , &  -\omega -  p_3\\      
    \end{pmatrix}\,,
    \label{eqGamma}
\end{equation}
so that the spectrum is given by  $\pm \epsilon_1$ and  $ \pm\epsilon_2$, with  
\[  \epsilon_1 = (p_1^2 - p_2^2)^{1/2} \,, \quad   \epsilon_2 =  p_3 \,. \]
We find that  $\epsilon_1\simeq \epsilon_2 \sim 1$ in the whole range of the fields, $b$, and these energies become equal at $b=b_c \simeq 0.6$. The corresponding energies of the Br and CCW modes at $\Gamma$ point are shown in Fig.\ \ref{fig:Gamma}. 
We see that the gap between the  Br and CCW modes closes at $b_c$ and re-opens at  further increase of $b$. 
In Fig.\ \ref{fig:Gamma} we also show the semi-quantitative agreement of our results with the full-scale calculation of the magnon spectrum, described in detail in Ref.\ \cite{Timofeev2022}. The main difference is the somewhat lower energy values for Br and CCW modes in full calculation, which can be associated with virtual transitions to higher energy levels, apparently present in full-scale calculation and  absent in the simplified formula \eqref{eq:L35}.   

%
%

{\bf Effective reduced Hamiltonian.}
To demonstrate the change in the topological character of two bands upon the gap re-opening, we focus 
on the vicinity of $\mathbf{q} =0 $. At $b=b_c$ and $\mathbf{q} =0$ the spectrum becomes doubly degenerate with $\omega=\pm \epsilon_1$. Our aim is first to reduce our analysis to a pair of identical secular equations, each given by 2$\times$2 Hamiltonian. 
There are two technical points arising here. First is the non-diagonal form of Eq.\ \eqref{eqGamma}, corresponding to finite hybridization between  skyrmion phase and radius even in the uniform limit, see definition of $u_{br}$ above. Such hybridization is absent in CCW mode at $\mathbf{q} =0$.  
Second, the kinetic term, $\hat{\mathcal K}_{\mathbf{q}}$, is non-diagonal, $k_{a,b} \neq0$, in contrast to the previously studied gyrotropic mode \cite{Timofeev2023a}, and this property turns out to be  important numerically. 

Exactly at the $\Gamma$ point we can bring our Lagrangian, $\hat{\mathcal L}_{\mathbf{q}}$, to the diagonal form by transformation 
$\hat{\mathcal L}_{\mathbf{q}=0} \to U_0^\dagger \, \hat{\mathcal L}_{\mathbf{q}=0} \,U_0 $, with 
\[ U_0 = \begin{pmatrix}
      \cosh \theta,& \sinh\theta ,& 0, & 0 \\ 
      \sinh  \theta,& \cosh \theta ,& 0, & 0 \\ 
    0, & 0,& -i,& 0 \\ 
    0, & 0,& 0,& i \\ 
\end{pmatrix}\,,\]
where $\tanh 2\theta = -p_2/p_1$. We consider now the  block of  $\hat{\mathcal L}_{\mathbf{q} }$, built on the first and third rows and columns.   The corresponding $2\times2$ block of kinetic term is unity matrix for  $\mathbf{q}=0$ and 
$      \begin{pmatrix}
      1+ k_2 \gamma_s   , &   \tilde{k}_a  \gamma_p ^\ast   , &     \\ 
             \tilde{k}_a  \gamma_p  , &    1+ k_3 \gamma_s ,   
      \end{pmatrix} $ 
otherwise, with $\tilde{k}_a$ defined below.  

\begin{figure}[t]
\center{\includegraphics[width=0.99\linewidth]{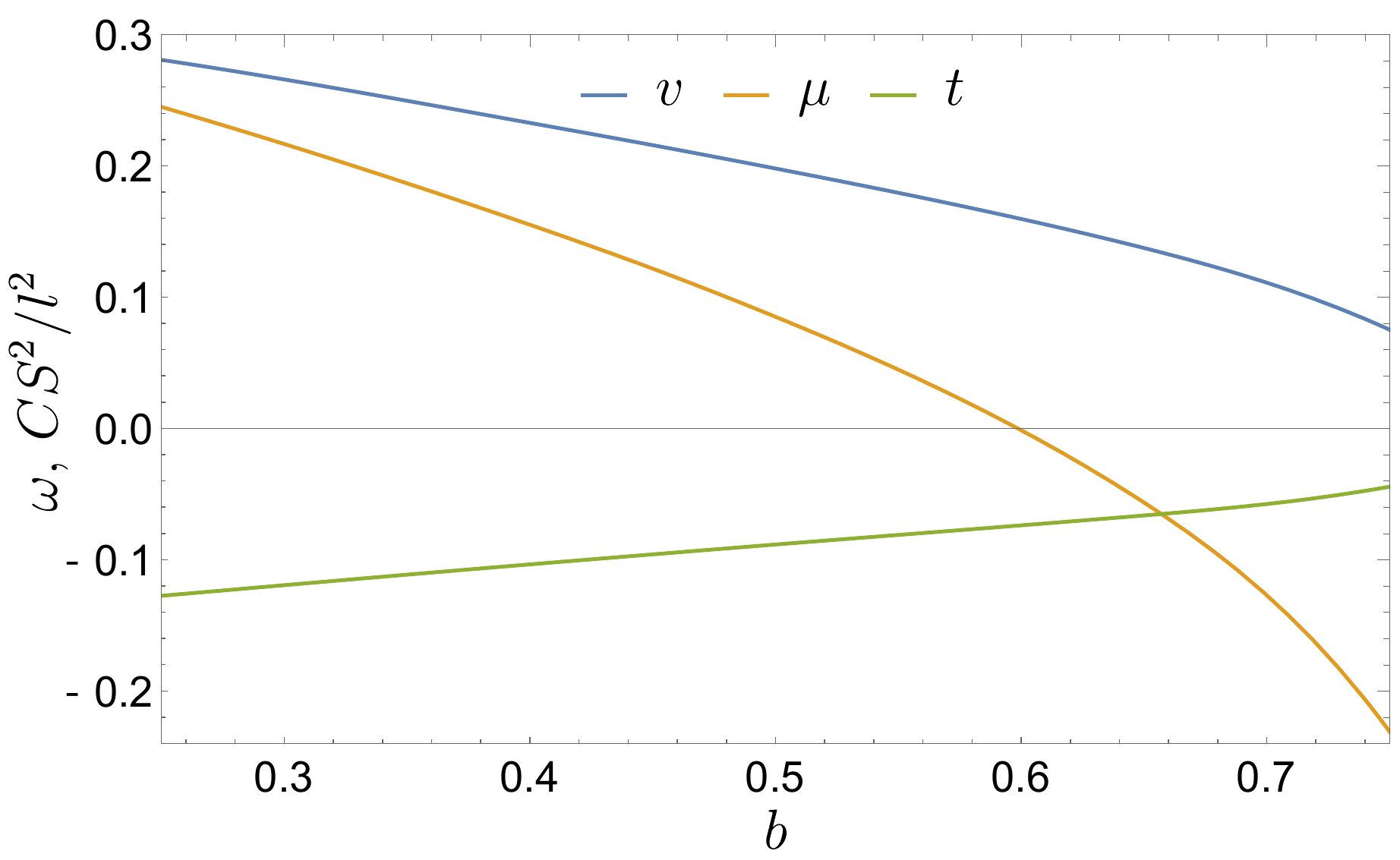}} 
\caption{ Parameters of the effective 2$\times$2 Hamiltonian, Eq. \eqref{eqHeff} as a function of the magnetic field, $b$.  }
\label{fig:Ham2x2}
\end{figure}

Proceeding our calculation to accuracy ${\cal O}(q^2)$, and applying simple $\mathbf{q}$-dependent similarity transformations to the Lagrangian, we can bring the above block of kinetic term to unity matrix again. 
Omitting further details, we present the  final expression
\begin{equation}     \begin{aligned} 
	\hat{\mathcal L}_{\mathbf{q} }^{(1)} & \simeq  (\omega - E_0) \mathbf{1} - {\cal H}  \,, \\ 
	{\cal H} & = 
	 \begin{pmatrix}
	\mu + t q^2 , & v(q_x-iq_y) \\  v(q_x+iq_y) , & -\mu - t q^2
	\end{pmatrix}    \,, \\ 
	\mu & =   ( \epsilon_1 - \epsilon_2)/2  , \quad
	 \epsilon_0  = (\epsilon_1 + \epsilon_2)/2\,, \\
	 v & = 3 (\tilde{t}_a + \epsilon_0\tilde{k}_a)    \,, \\ 
	t &  =  \tfrac34 ( t_3-  \tilde{t}_1 +k_2 \epsilon_1 -  k_3 \epsilon_2  + 6 \tilde{k}_a^2 \mu     ) \,, \\  
	 E_0 & = \epsilon_0 +  \tfrac34 q^2 
	 ( -t_3 - \tilde{t}_1 +k_2 \epsilon_1+   k_3 \epsilon_2   \\ &   
           + 12 \tilde{k}_a \tilde{t}_a + 12 \tilde{k}_a^2  \epsilon_0  )
	  \,, 
	  \label{eqHeff}
\end{aligned} \end{equation}
where 
\begin{equation}     \begin{aligned} 
	 \tilde{t}_a & = t_a \cosh \theta + t_b \sinh \theta  \,, \\
	\tilde{k}_a & = k_a \cosh \theta + k_b \sinh \theta  \,, \\ 
	\tilde{t}_1 &= t_1 \cosh 2\theta - t_2 \sinh2\theta 
	  \,.  
\end{aligned} \end{equation}
 The second block, $\hat{\mathcal L}_{\mathbf{q} }^{(2)}$, is obtained  from  $\hat{\mathcal L}_{\mathbf{q} }^{(1)}$ by changing $\omega\to -\omega$ and $q_y\to - q_y$. 
 Two remaining blocks in $\hat{\mathcal L}_{\mathbf{q} }$, connecting $\hat{\mathcal L}_{\mathbf{q} }^{(1)}$ and $\hat{\mathcal L}_{\mathbf{q} }^{(2)}$, are small, ${\cal O}(q)$, and can be discarded near $\Gamma$ point.  
The dependence of the parameters of the effective Hamiltonian $ {\cal H} $ on the  field $b$ is shown in Fig.\ \ref{fig:Ham2x2}. 
 The eigenmodes of the Hamiltonian \eqref{eqHeff} are given by  
\begin{equation} 
	\epsilon_{\pm,\mathbf{q} } = E_0 \pm \sqrt{ (\mu + t q^2) ^2+v^2q^2 }\,.
\end{equation}

The topological character of the spectrum is characterized by  the Berry curvature, $\Omega(\mathbf{q} )$, which is readily evaluated 
for our simplified model, Eq. \eqref{eqHeff}. According to the usual formula, if the $n$th band is characterized by dispersion $\epsilon_{n \mathbf{q}} $, then the Berry curvature for this band, $\Omega_n(\mathbf{q} )$, is given by
\begin{equation}
\Omega_n(\mathbf{q}) = i \sum_m 
\frac{ \langle n | \frac{\partial {\hat H}_\mathbf{q} }{\partial q_x} | m\rangle  
\langle m | \frac{\partial {\hat H}_\mathbf{q}}{ \partial q_y} | n\rangle - (q_x \leftrightarrow q_y)}
{ (\epsilon_{n \mathbf{q}} - \epsilon_{m \mathbf{q}})^2} \,.
\label{BerryCurv0}
\end{equation}
Generally, the summation over $m$ should include all bands of the spectrum  in \eqref{BerryCurv0}. 
In our case of two closely situated levels, $\epsilon_{+,\mathbf{q} } - \epsilon_{-,\mathbf{q}} \ll E_0\sim 1$, we can consider only transitions  between these levels, when the denominator in  \eqref{BerryCurv0} is small. The transitions with $|\epsilon_{n \mathbf{q}} - \epsilon_{m \mathbf{q}}| \simeq 2E_0$ may be discarded.    It means that the sum over $m$ reduces to one term only. 
Some calculation yields the expression for the band $\epsilon_{+,\mathbf{q} } $
\begin{equation}
\Omega(\mathbf{q} ) = \frac{v^2(  t q^2-\mu) }{2 \left( (\mu + t q^2) ^2+v^2q^2 \right)^{3/2}}\,. 
\label{Oq2x2}
\end{equation}
and one has to replace $\Omega(\mathbf{q} )\to - \Omega(\mathbf{q} )$ for  $\epsilon_{-,\mathbf{q} } $.

The integrated weight (Chern number) is given by 
\[
\begin{aligned}
\int d^2\mathbf{q} \, \Omega (\mathbf{q} ) & =
 \frac{\pi (  t q^2+\mu) }   { \sqrt{ (\mu + t q^2) ^2+v^2q^2 }} \big|_{q=0}^\infty  
 \\ & = 
  \pi  (\mbox{sgn}\,t  -  \mbox{sgn}\,\mu)\,.
\end{aligned}
\]
We see that for $\mu t < 0$ the magnons are topologically non-trivial, which happens in our case at $b<b_c \simeq 0.6$. 
For larger fields, $b>b_c$,  we have $\mu t >0$, so that  the spectrum becomes  topologically trivial. 
Such a transition was observed earlier by increasing magnetic field in a similar model \cite{Diaz2020} or magnetic anisotropy strength in model lattice system with four-spin interaction \cite{Maeland2022}. 

\begin{figure}[t]
\center{\includegraphics[width=0.99\linewidth]{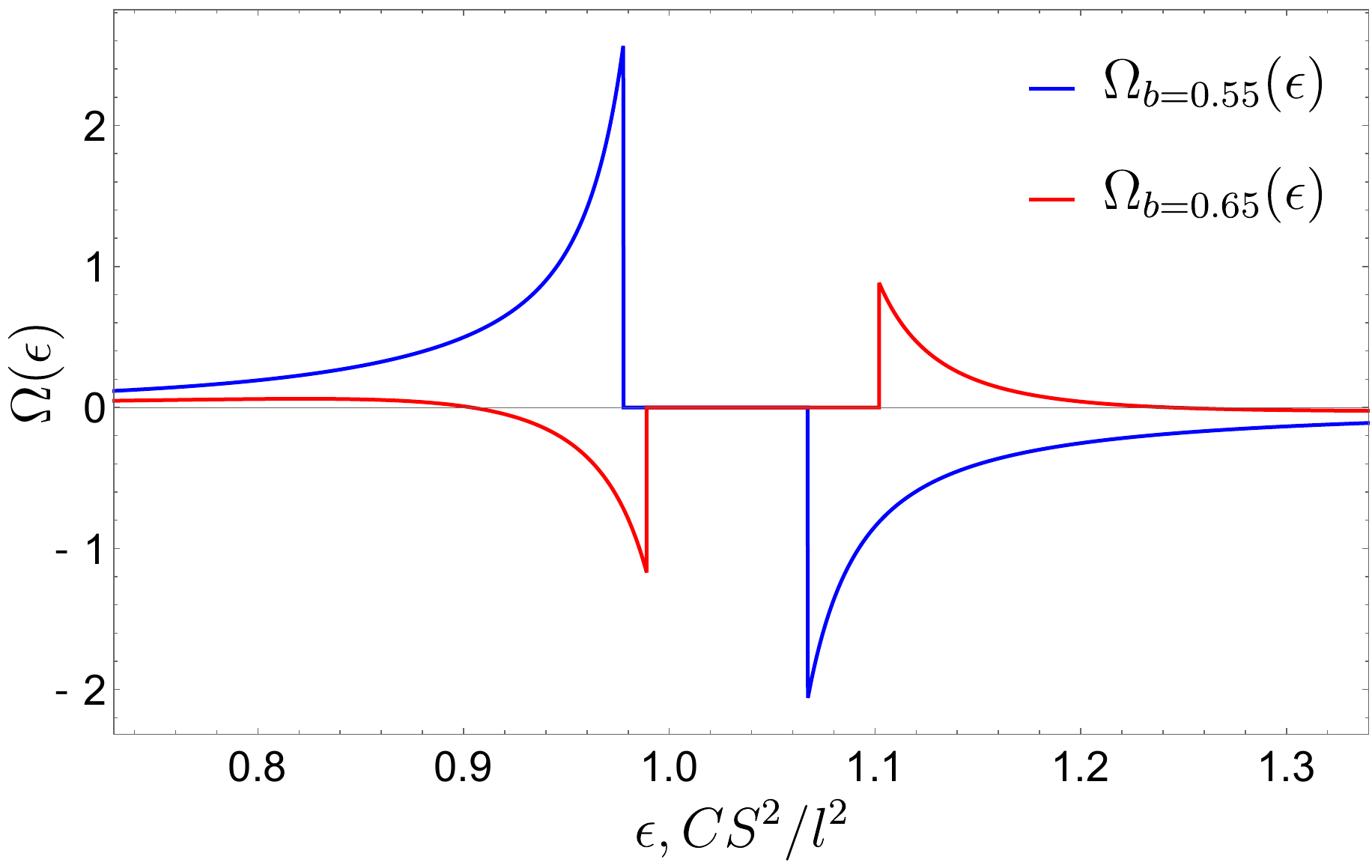}} 
\caption{ Density of Berry curvature $\Omega(\epsilon)$ for two different field values: the blue line $b=0.55$ and the red line $b=0.65$.}
\label{fig:Omega}
\end{figure}

The topological character of the magnon spectrum can manifest itself in appearance of edge states \cite{roldan2016,Diaz2020} and in anomalies of thermal Hall conductivity, see Ref.~\cite{PhysRevB.89.054420}.
In the latter application of the discussed model it may be 
 convenient  to use the density of Berry curvature, which we define as  
\begin{equation}
\Omega(\epsilon) =  V^{-1} \sum_{n \mathbf{q}} \Omega_n(\mathbf{q}) \delta(\epsilon - \epsilon_{n \mathbf{q}})\,.
\label{DoBC}
\end{equation}
In the vicinity of the gap reopening at $b=b_c$ this function for our simplified model  \eqref{eqHeff} is reduced to 
\begin{equation}
\Omega(\epsilon) =    \int \frac{d^2\mathbf{q}}{(2\pi)^2} 
\Omega(\mathbf{q} ) \left( \delta(\epsilon - \epsilon_{+,\mathbf{q} } ) -  \delta(\epsilon - \epsilon_{-,\mathbf{q}} ) \right)\,,
\label{DoBC2x2}
\end{equation}
with $\Omega(\mathbf{q} )$ given by \eqref{Oq2x2}. The appearing expressions are quite complicated and not shown here. 
 Fig. \ref{fig:Omega} shows the behavior of   the function $\Omega(\epsilon)$ for actual set of parameters $ E_0$,  $v$, $\mu$,  $t$ in two cases, corresponding to    $\mu t >0$ and $\mu t < 0$.
  
%
%

{\bf Conclusions.} Summarizing, we develop a theory of the topological transition happening in the hexagonal skyrmion crystal upon the increase of the external magnetic field. The energies of two low-lying magnon modes, visible in magnetic resonance experiments, intersect each other upon this increase, which is accompanied by the change in the topological character of both magnon bands. We hope that this phenomenon, if confirmed by means of magnetic resonance, can be further investigated in thermal Hall conductivity experiments.

{\bf Acknowledgements}. 
The work was supported by the Russian Science Foundation, Grant No. 22-22-20034 and St.Petersburg Science Foundation, Grant No. 33/2022. 

\bibliography{skyrmionbib}

\end{document}